\documentclass[prd,10pt,aps,nofootinbib, floatfix, notitlepage,longbibliography,superscriptaddress,preprintnumbers]{revtex4-2}

\usepackage{placeins}
\usepackage{amsmath,amsfonts,amssymb}
\usepackage{mathrsfs}
\usepackage{graphicx}
\usepackage[english]{babel} 
\usepackage{ulem}
\usepackage{url} 
\usepackage{color}
\usepackage{bbold}
\usepackage{adjustbox}
\usepackage[utf8]{inputenc} 
\usepackage{mathtools}
\usepackage[colorlinks=true,citecolor=blue,urlcolor=blue]{hyperref}

\begin{document}
	
	\title{Improved hot dark matter bound on the QCD axion}

    \author{Alessio~Notari}
	\email{notari@fqa.ub.edu}
	\affiliation{Departament de F\'isica Qu\`antica i Astrofis\'ica \& Institut de Ci\`encies del Cosmos (ICCUB), Universitat de Barcelona, Mart\'i i Franqu\`es 1, 08028 Barcelona, Spain}
 \affiliation{Galileo Galilei Institute for theoretical physics, Centro Nazionale INFN di Studi Avanzati
Largo Enrico Fermi 2, I-50125, Firenze, Italy
		\looseness=-1}
	\author{Fabrizio Rompineve}
	\email{fabrizio.rompineve@cern.ch}
	\affiliation{CERN, Theoretical Physics Department, 1211 Geneva 23, Switzerland
		\looseness=-1}
	\author{Giovanni Villadoro}
	\email{giovanni.villadoro@ictp.it}
	\affiliation{Abdus Salam International Centre for Theoretical Physics, Strada Costiera 11, 34151, Trieste, Italy
		\looseness=-1}
	
	\date{\today}
	
	\preprint{CERN-TH-2022-165}
	
\begin{abstract}
	
	\noindent 
 We strengthen the cosmological bound on the axion mass, by solving the momentum-dependent
Boltzmann equations for axion-pion
scatterings and by using a phenomenological production rate derived from pion-pion scattering data, overcoming the breakdown of chiral perturbation theory. Using present cosmological datasets we obtain $m_a\leq 0.24~\text{eV}$. 
To further improve the bound and exploit the reach of upcoming cosmological surveys, reliable non-perturbative calculations above the QCD crossover are needed.

	\end{abstract}

	\maketitle

\section{Introduction} 

The Peccei-Quinn solution to the strong CP problem~\cite{Peccei:1977hh, Peccei:1977ur} involves an extra field \cite{Weinberg:1977ma, Wilczek:1977pj}, the axion $a$, coupled to QCD. In the cosmic evolution scatterings mediated by such couplings produce a background of ``hot" relic axions~\cite{Turner:1986tb, Kolb:1990vq} (see also refs.~\cite{Berezhiani:1992rk, Chang:1993gm, Masso:2002np} for early follow-ups), very much analogous to the neutrino background and distinct from cold axions produced via the misalignment mechanism~\cite{Abbott:1982af,Dine:1982ah,Preskill:1982cy} and the decay of topological defects~\cite{Davis:1986xc, Harari:1987ht, Lyth:1991bb}. 

Available cosmological datasets~\cite{Planck:2018vyg} already constrain light relics produced at temperatures slightly below the QCD crossover, i.e. $T\lesssim T_c\equiv 150~\text{MeV}$, while important sensitivity improvements of current~\cite{ACT:2020gnv, SPT-3G:2021vps, DESI:2016fyo} and upcoming~\cite{SimonsObservatory:2018koc, CMB-S4:2022ght, Amendola:2016saw, LSST:2008ijt} Cosmic Microwave Background (CMB) and Large Scale Structure (LSS) surveys will allow to probe production during the crossover. A QCD axion with mass $m_a\sim (0.01-1)~\text{eV}$ would be produced precisely around these temperatures, and can thus be constrained/discovered with cosmology, independently of astrophysical probes (for SN1987A, see ref.~\cite{Raffelt:1990yz} and refs.~\cite{Chang:2018rso, Bar:2019ifz, Carenza:2020cis} for recent conflicting reassessments; for the most recent constraints from other sources see refs.~\cite{MillerBertolami:2014rka, Ayala:2014pea, Dolan:2022kul, Buschmann:2021juv}), in a region targeted also by direct searches~\cite{IAXO:2019mpb, IAXO:2020wwp}. Extracting a reliable prediction for the relic abundance of such axions is thus of crucial observational relevance.

In this Letter, we significantly improve on previous such predictions in two aspects. First, the rapid and significant variation of the number of relativistic species $g_{*, S}$ around $T_c$ implies that the axion spectrum can significantly deviate from the equilibrium distribution that has been employed so far. We account for this by solving the momentum-dependent Boltzmann equations for axion scatterings. This strategy is used for the relic neutrino abundance~\cite{Hannestad:1995rs, Dolgov:1997mb}, which however occurs when the variation of $g_{*,S}$ is much less relevant, thus causing a smaller distortion (see ref.~\cite{Dunsky:2022uoq} for a related work focused on heavy axions). Second, the axion production rate of interest for current datasets, dominantly due to scatterings with pions ($\pi\pi\leftrightarrow a\pi$), has been historically computed at leading order (LO) in chiral perturbation theory ($\chi$PT)~\cite{Chang:1993gm} (see refs.~\cite{Hannestad:2005df, Melchiorri:2007cd, Hannestad:2007dd, DiValentino:2015wba} for the corresponding ``hot dark matter" bound on $m_a$ from CMB data). Nonetheless, recent work~\cite{DiLuzio:2021vjd} finds that such a rate receives large one-loop corrections already at $T\gtrsim 60$  MeV, casting doubts on the validity of existing constraints. We overcome  this obstacle by deriving the $a$-$\pi$ rate from the experimental data on $\pi\pi\leftrightarrow \pi\pi$ scattering, via a  simple rescaling of the corresponding cross sections, as dictated by the mixing between the axion and the pion. 
These two improvements allow us to place a robust upper bound on the axion mass using cosmological datasets. 

Finally, we highlight previously neglected non-perturbative contributions at $T_c$ and above, which hinder the reliability of perturbative approaches and call for a dedicated study to further improve the present bound and to correctly interpret the phenomenological implications of forthcoming experiments.

\section{Boltzmann equations} 

We focus on the minimal model-independent axion interaction, ${\cal L}_{\rm int}=\alpha_s a G\tilde{G}/(8\pi f_a)$, where $G$ is the gluon field strength, with its dual $\tilde{G}$, $\alpha_s$ is the strong coupling constant and $f_a$ is the axion decay constant, related to the axion mass by $m_a=0.57~\text{eV}(10^7~\text{GeV}/f_a)$~\cite{Gorghetto:2018ocs}. 

An important novelty of our work is the use of the momentum-dependent Boltzmann equations to compute the actual spectrum of axions produced via scatterings. We find this to be necessary, for two reasons. First, the initial axion abundance may be negligible and production may always remain sub-thermal, such that the axion spectrum may always significantly differ from the equilibrium distribution. Second, even when reaching equilibrium, interaction rates depend on the axion momentum, so that high momenta decouple later than low momenta. Since $g_{*, S}$ decreases quickly and substantially with temperature around~$T_c$, high momenta are less diluted with respect to photons than low momenta.

In the expanding Universe it is convenient to use comoving 3-momenta ${\bf p}$, such that the axion distribution function $f_{{\bf p}}$ remains constant in the absence of interactions. The Boltzmann equations then reads~\cite{Kolb:1990vq, Bellac:2011kqa}:
\begin{eqnarray}
\label{eq:boltzmann}
\frac{df_{{\bf p}}}{dt} = (1+ f_{{\bf p}})\, \Gamma^< - f_{{\bf p}}\, \Gamma^>\, , 
\end{eqnarray}
where the axion, with negligible mass at the time of production, has physical energy $E=|{\bf p}|/R$ and four momentum $k^\mu=(E,{\bf k})$, $R$ is the scale factor, $t$ is cosmic time and the energy dependent rates $\Gamma^<$ and $\Gamma^>$ describe creation and destruction of an axion. Assuming thermal equilibrium of the QCD bath, the following non-perturbative relation holds:
		\begin{eqnarray} \label{eq:axionrate}
	\label{gammas}
	\Gamma^>=e^{\frac{E}{T}} \Gamma^< =\frac{\Gamma^>_{\rm top}}{2E f_a^2}\,, \qquad 
	\Gamma^>_{\rm top}\equiv \int\hspace{-4pt}d^4x\,e^{ik^\mu x_\mu}\,\left\langle \frac{\alpha_s}{8\pi}G\tilde G(x^\mu)\,\frac{\alpha_s}{8\pi}G\tilde G(0)\right\rangle\,, 
		\end{eqnarray}
where $\langle \cdots \rangle$ stands for thermal average. 
At weak coupling the rates are dominated by $2\leftrightarrow 2$ scatterings, with amplitude $\mathcal{M}$, 
leading to~\cite{Bellac:2011kqa}:
\begin{eqnarray} \label{eq:Gamma<}
 \Gamma^<   =  \frac{1}{2 E}\int \left(\prod_{i=1}^3\frac{d^3{\bf k}_{i}}{(2\pi)^3 2E_{i}}\right)
 f^{\rm eq}_{{1}} f^{\rm eq}_{{2}}  (1+f^{\rm eq}_{{3}}) (2\pi)^4\delta^{(4)}(k^\mu_{1}+k^\mu_{2}-k^\mu_{3}-k^\mu)|\mathcal{M}|^2  \,,
		\end{eqnarray}
where $f^{\rm eq}_{i}\equiv (e^{{E_i}/T} \mp 1)^{-1}$ is the equilibrium distribution for bosons and fermions. For instance, for axion-pion scatterings, $\pi\pi\leftrightarrow \pi a$, the physical four momenta are $k_{1}$ and $k_{2}$ for the incoming pions, $k_{3}$ for the emitted pion  (gluons or other particles can be treated analogously).

The axion relic abundance is then commonly expressed in terms of the effective number of (massless) neutrino species beyond the SM neutrinos $N_\nu\approx 3.044$~\cite{Froustey:2020mcq, Bennett:2020zkv}:
\begin{equation}
\label{eq:deltaneff}
    \Delta N_{\text{eff}}\equiv N_{\text{eff}}-N_\nu = \frac{8}{7}\left(\frac{11}{4}\right)^{\frac{4}{3}}\left(\frac{\rho_a}{\rho_\gamma}\right)_{\text{CMB}},
\end{equation}
where the energy density of axions, $\rho_a$, and photons, $\rho_\gamma$, should be evaluated at recombination, when the Cosmic Microwave Background (CMB) is generated. In the following we shall make contact with previous approaches in the literature, which use a momentum-independent Boltzmann equation (see Appendix A), through the averaged rate
\begin{equation}
\label{eq:totrate}
    \overline{\Gamma}\equiv \frac{1}{n^{\text{eq}}}\int \frac{d^3{\bf k}}{(2\pi)^3}\Gamma^<,
\end{equation} 
with $n^{\text{eq}}\equiv\int d^3{\bf k}/(2\pi)^3 f_{\bf p}^{\text{eq}}$. Furthermore, under the assumption of instantaneous decoupling of an initial equilibrium population of axions, eq.~\eqref{eq:deltaneff} is approximated as $\Delta N_{\text{eff}}\simeq~0.027\left[106.75/g_{*, S}(T_{\text{d}})\right]^{4/3}$, with $T_d$  defined by $\overline{\Gamma}=H\rvert_{T=T_d}$ and $H\equiv \dot{R}/R$. As mentioned above, neither of these two approaches is justified in our case.

In our region of interest $m_a$ is actually close to the temperature at recombination ($\lesssim \text{eV}$). $\Delta N_{\text{eff}}$ should then be interpreted as the would-be abundance of axions, if they were massless. As we will show, the bound on the axion contribution to $\Delta N_{\text{eff}}$ will be significantly stronger than the massless species constraint, $\Delta N_{\text{eff}}\leq 0.28$ at $95\%$~C.L. (Planck 18+BAO)~\cite{Planck:2018vyg}.

\begin{figure*}[t]
		\includegraphics[width=0.48\textwidth]{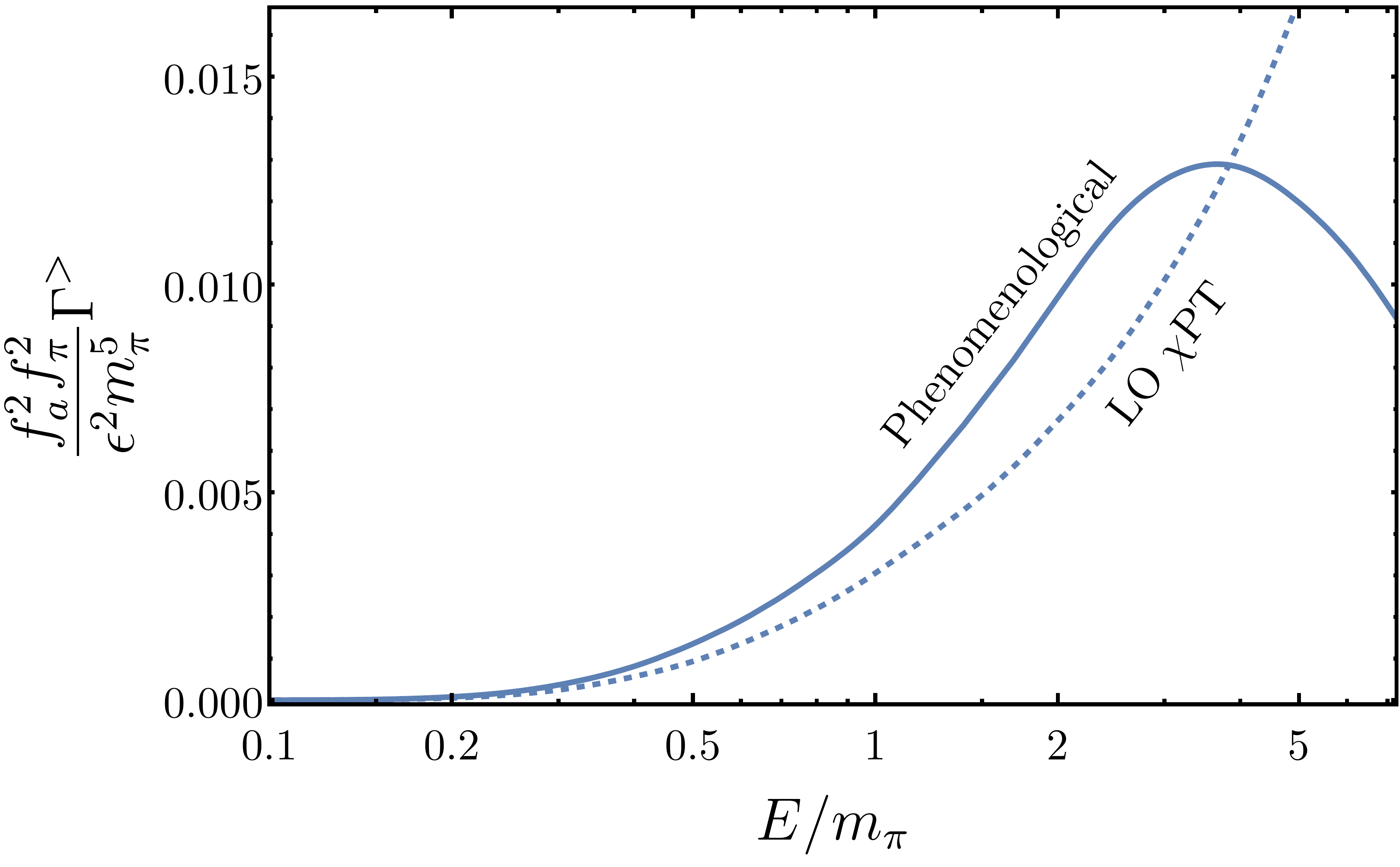}
         \hspace{0.5cm}
	    \includegraphics[width=0.47\textwidth]{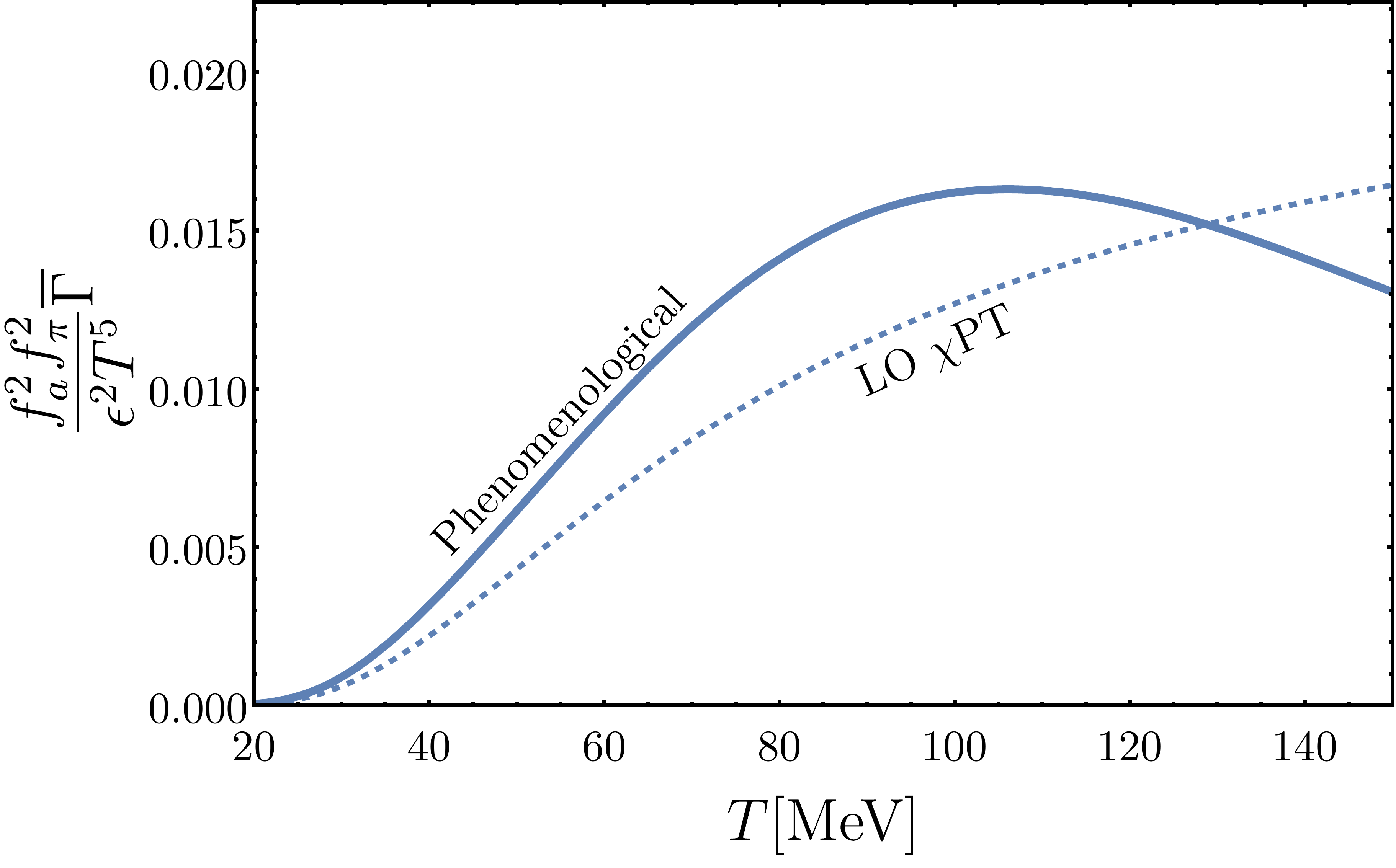}
		\caption{Scattering rates for $a\pi \leftrightarrow \pi\pi$ (solid curves for phenomenological rates, dashed for LO in $\chi$PT). {\it Left:} momentum dependent axion destruction rates  at $T=120$ MeV. {\it Right:} Averaged rates.}
		\label{fig:axpion}
\end{figure*}

\section{Axion rate below $T_c$}
\label{pionrate}

At $T\lesssim T_c$ the QCD thermal bath is dominated by pions, all other hadrons being Boltzmann suppressed to some extent.  As already stressed, recently  ref.~\cite{DiLuzio:2021vjd} pointed out that next-to-leading-order (NLO) corrections 
invalidate the LO $\chi$PT computation, already at $T\simeq 60~\text{MeV}$. In retrospect this is not surprising given that
1) the typical center of mass energy $\sqrt{s}$ for a pair of pions at such temperatures is already above 0.4~GeV and 2) the scattering amplitudes grow with
energy, thereby pions with even higher energies, for which the $\chi$PT estimate is unreliable, are weighted more in the integral (\ref{eq:totrate}).

In fact the very same problem was encountered long ago for the pion damping rates \cite{Goity:1989gs, Bebie:1991ij, Schenk:1993ru}, and it was circumvented by using experimental $\pi$-$\pi$ scattering data directly, rather than relying on 
$\chi$PT. With this strategy, the pion rate was computed up to $T=T_c$ (beyond which the rate becomes rapidly
comparable with the mass and pions cannot be considered as elementary particles anymore). 
In an attempt to rescue the $\chi$PT computation ref.~\cite{Schenk:1993ru} proposed also a unitarization approach. 
The idea is to apply the $\chi$PT expansion to (functions of) the scattering phase shifts 
rather than to the amplitude $\mathcal{M}$ itself. In this way unitarity is always respected
and the growth at high energies tamed.
The unitarization procedure is however not unique. 
While some choices agree well with experimental data, without the latter it would be hard
to defend the use of one particular prescription over the others, 
or why it would be a good approximation to neglect even higher-order corrections.

Unfortunately, for axions there are no experimental data and at a first glance it seems that only unitarization could improve analytically on the fixed order $\chi$PT computation. 
However, we prove (see Appendix~\ref{app:apiscattering}) the following relation between $a$-$\pi$ and
$\pi$-$\pi$ scattering amplitudes at all orders in $\chi$PT 
\begin{equation} \label{eq:MapfromMpp}
\mathcal{M}_{a \pi^i\to \pi^j \pi^k}=\frac{\epsilon \,f_\pi}{2f_a}\cdot \mathcal{M}_{\pi^0 \pi^i\to \pi^j \pi^k}+{\cal O}\left(\frac{m_\pi^2}{s} \right)\,, \qquad \epsilon\equiv \frac{m_d-m_u}{m_d+m_u}\,,
\end{equation}
where $f_\pi=92.3~\text{MeV}, m_\pi=138~\text{MeV}$ (we use the average pion mass), $m_u$ and $m_d$ are up and down quark masses, $m_u/m_d\simeq 0.47$~\cite{ParticleDataGroup:2022pth}.
We checked this explicitly at LO and NLO by comparing ref.~\cite{Gasser:1983yg} for $\pi$-$\pi$ and refs.\cite{Chang:1993gm,DiLuzio:2021vjd} for $a$-$\pi$ scattering amplitudes. 
The ${\cal O}(m_\pi^2/s)$ corrections near the two pions threshold can be computed directly at LO (see Appendix~\ref{app:apiscattering}) and they are at most ${\cal O}(10\%)$  and rapidly decrease at higher energies. While here we focused on the model-independent coupling of the QCD axion, eq.~(\ref{eq:MapfromMpp}) can easily be extended to accommodate the most general axion couplings, see appendix~\ref{app:apiscattering}.

Thanks to eq.~\eqref{eq:MapfromMpp} phenomenological fits of $\pi$-$\pi$ scattering data can be used to reconstruct $a$-$\pi$ scattering amplitudes with 
a few percent precision up to $T_c$.
To compute the momentum-dependent rate we applied eq.~(\ref{eq:MapfromMpp}) to the phenomenological $\pi$-$\pi$ partial wave amplitudes provided in refs.~\cite{Pelaez:2004vs, Garcia-Martin:2011iqs} (specifically the S0, S2 and P waves, valid up to the two-kaon threshold $\sqrt{s}=2m_K\simeq 1~\text{GeV}$, while we checked that higher partial waves give negligible contributions). The result is presented in Fig.~\ref{fig:axpion} (solid blue curve, left panel), for a reference temperature $T=120~\text{MeV}$. As expected, it decreases sharply at large momenta, in contrast to the LO $a$-$\pi$ rate (dashed). The corresponding averaged rates are shown in Fig.~\ref{fig:axpion} (right panel). 

The decrease of the $a$-$\pi$ rate at high temperatures is in part due to the opening of new scattering channels (see~\cite{Bebie:1991ij} for an analogous discussion for pions). Using the LO $\chi$PT Lagrangian, we checked for instance that scatterings with Kaons $\pi K\rightarrow a K$ are subleading below $T_c$, but would eventually dominate above $T\simeq 200~\text{MeV}$, see Appendix~\ref{app:crosssections} (similar considerations may apply to scattering off nucleons). While in general a $\chi$PT calculation for these processes is less reliable, such an estimate shows that the  phenomenological $a$-$\pi$ rate provides the most important contribution for the axion thermalization rate at low temperatures, representing a reliable lower bound at $T\lesssim T_c$. 

\begin{figure*}[t]
		\includegraphics[width=0.8\textwidth]{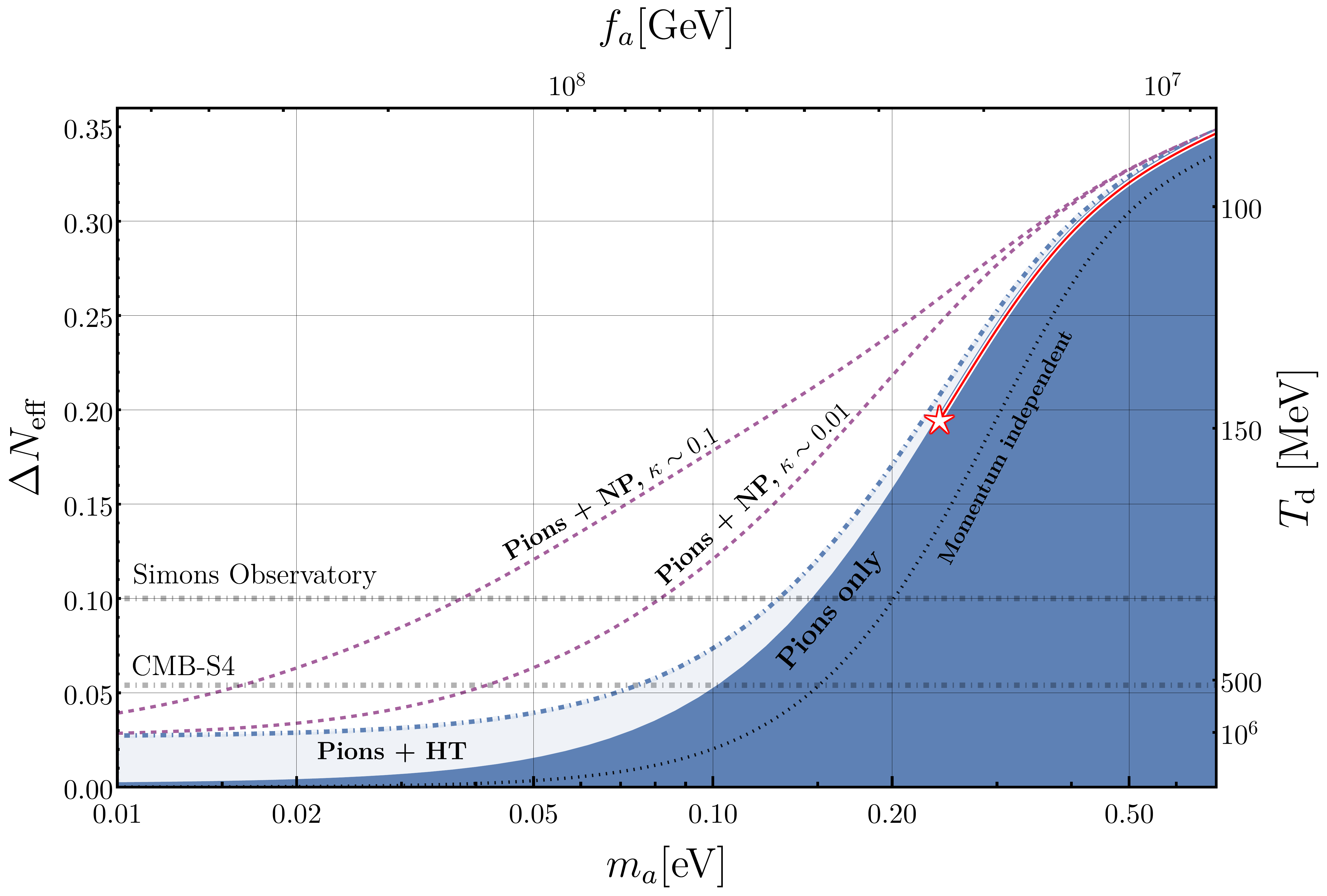}

		\caption{Relic axion abundance parameterized by $\Delta N_{\rm eff}$, for an axion minimally coupled to QCD. The blue region is our lower bound on $\Delta N_{\rm eff}$ obtained by using the full momentum-dependent Boltzmann equations with phenomenological $\pi\pi\leftrightarrow a\pi$ rate below $T_c$ only, and with zero initial conditions at $T_c$. For comparison, results with the ``standard"  momentum-independent Boltzmann equation (dotted black) or assuming an equilibrium population of axions above $T=1~\text{TeV}$ (``Pions+HT", dot-dashed) are shown. Tentative dimensional analysis estimates for non-perturbative production up to $T=2~\text{GeV}$ are plotted as dashed purple curves (``Pions+NP"). On the right vertical axis, the temperature $T_d$ corresponding to axions produced with instantaneous decoupling is plotted. The red starred point corresponds to the $95\%$~C.L. bound that we obtain using CMB, BAO and Pantheon data. The $95\%$~C.L. expected sensitivities of upcoming CMB surveys for massless species are shown by the dot-dashed gray lines.}
		\label{fig:DeltaNeff}
	\end{figure*}

Integrating numerically the momentum dependent Boltzmann equations (\ref{eq:boltzmann}) (see Appendix~\ref{app:boltzmann}) and assuming conservatively no extra production 
from $T>T_c$, we get the lower bound on $\Delta N_{\rm eff}$ represented by the boundary of the solid
blue region in fig.~\ref{fig:DeltaNeff}. Because of the rapid change in $g_{*, S}$, during the axion
decoupling, the resulting spectrum is distorted and $\rho_a$ is enhanced compared to the standard calculation without momentum dependence (dotted black curve). A slightly less conservative assumption is to consider the axion in thermal equilibrium at some temperature $T>1$~TeV (with $g_{*, S}$ saturated by the Standard Model). For this case, the bound is represented by the boundary of the light blue region in fig.~\ref{fig:DeltaNeff}. 

Let us also mention that the axion coupling to pions can be slightly enhanced or suppressed with respect to the value considered here (i.e. corresponding to so-called hadronic models~\cite{Kim:1979if, Shifman:1979if}), in scenarios where the QCD axion also couples to quarks at high energies (e.g.~\cite{Zhitnitsky:1980tq, Dine:1981rt}, see refs.~\cite{Baumann:2016wac,Ferreira:2018vjj,DEramo:2018vss,AriasAragon:2020shv,Green:2021hjh,Ferreira:2020bpb} for relic abundance calculations). We do not consider these here, since they are generically more constrained by astrophysical observations~\cite{MillerBertolami:2014rka} and the resulting $\Delta N_{\text{eff}}$ is necessarily model-dependent.  

\section{Estimates above $T_c$}

While thermalization rates at the QCD confining scale are obviously non-perturbative, naively one would expect 
that at temperatures above a few~GeV perturbative QCD should provide reliable 
rates. After all,
at those temperatures the typical momentum of quarks and gluons is definitely in the perturbative regime.
However, it has long been known that at finite temperature the convergence of perturbative QCD is worsened by various infrared divergences \cite{Linde:1980ts,Gross:1980br}. One of them is manifest already at the leading order for the case at hand. 
The leading processes involve gluons ($g$) and quarks ($q$): $a g\leftrightarrow g g$,  $a q\leftrightarrow  g q$ and  $a g\leftrightarrow  q \bar{q}$. The gluon exchange in the $t$ channel leads to an infrared divergence~\cite{Masso:2002np, Graf:2010tv, Salvio:2013iaa} for the first two processes. At weak coupling ($g_s=\sqrt{4\pi \alpha_s}\ll 1$) 
the divergence is regulated by the gluon screening mass $m_g\sim g_s T$, leading to a logarithmic, but finite, 
enhancement in the rate proportional to $\log(T/m_g)$. However, $g_s$ is still larger than unity
at the electro-weak scale and only decreases logarithmically at higher energies. This makes the infrared logarithm and
the total rate turn unphysically negative, unless the temperature is exponentially large. 

This issue has been improved in ref.~\cite{Salvio:2013iaa} (see also ref.~\cite{Rychkov:2007uq}) using the full resummed thermal gluon propagator computed at LO in $g_s$. Such an approach however is
not fully consistent, since vertex corrections are not properly resummed 
and gauge invariance not completely restored at the order considered. 
While ref.~\cite{Salvio:2013iaa} used this method for temperatures $T\gtrsim 10^4~\text{GeV}$, where $g_s\lesssim 1$, ref.~\cite{DEramo:2021lgb} pushed it all the way down to $T=2$~GeV, where $g_s\simeq 2$.

It should be pointed out however that perturbation theory can really be trusted only for
$g_s\ll 1$. One way to see this is to consider the ratio of the gluon thermal width over its electric mass, $\Gamma_g/m_g$. When this becomes ${\cal O}(1)$ the gluon is not a good weakly coupled
degree of freedom. At  leading order (see e.g.~\cite{Bellac:2011kqa})
$\Gamma_g/m_g\sim 0.5 g_s$,  implying that perturbativity is really under control only at $T\gg$~TeV.
Another requirement for the case at hand is the existence of a clear separation between $T$, the typical momentum scale of the thermal bath, and $g_s^2 T$, the momentum scale below which collective behaviors become important and gluon interactions become non perturbative. As long
as $g_s\ll 1$ such soft momentum region is phase space suppressed, so that the perturbative expansion
is meaningful up to a certain finite order \cite{Gross:1980br}. 
When $g_s\gtrsim 1$
non-perturbative effects may become dominant and perturbation theory unreliable.  

Accidentally, the non-perturbative axion  damping rate  at zero momentum has been studied in a different context. Such a quantity is indeed related via eq.~(\ref{eq:axionrate}) to the so-called strong-sphaleron rate $\Gamma_{\text{sph}}\equiv \Gamma^>_{\rm top}(k^\mu=0)$ \cite{McLerran:1990de}. While non-perturbative in nature, at small enough couplings
its contribution has been inferred \cite{Bodeker:1998hm,Arnold:1998cy} and checked numerically using classical field simulations \cite{Moore:2010jd}, leading to the approximate scaling 
$\Gamma_{\text{sph}}\simeq (N_c \alpha_s)^5 T^4$.  If extrapolated at small energies this
can become comparable to the putative perturbative computation at temperatures well above GeV (see Appendix~\ref{app:sphaleron} for more details). 

We therefore conclude that axion-gluons  perturbative calculations cannot
be trusted below the electro-weak scale. In particular below the GeV scale, the most interesting region of temperatures for upcoming CMB experiments,
the rate is definitely non-perturbative. Besides, between 150~MeV and 500~MeV the entropy and the energy densities have the largest variation
associated to the QCD crossover~\cite{Borsanyi:2016ksw}. This will enhance the effects of the spectral distortion possibly even more than
that observed in our pion computation; it becomes therefore mandatory to study directly
$\Gamma^{>}_{\rm top}(E)$ of eq.~(\ref{eq:axionrate}) with non-perturbative methods to further improve the bound and correctly assess the implications of forthcoming experiments on the axion parameter space.

At present we are not aware of any such computations in this regime, a challenging task beyond the aim of this work. 
All we can do is to attempt a very rough estimate of the averaged rate during the crossover, 
say for $\, T_c\lesssim T \lesssim 2 \, {\rm GeV}$, using dimensional analysis, i.e. $\overline{\Gamma}\sim \kappa \, T^3/f_a^2$, 
where $\kappa$ is an unknown coefficient. For illustrative purposes we choose two reference values 
$\kappa=0.01$ and $\kappa=0.1$, justified as follows.

First, at $T\gtrsim f_\pi$ the $a$-$\pi$ rate just below $T_c$ is $\sim 0.04~\epsilon^2 T^3/f_a^2$ (see also Fig.~\ref{fig:tentrate} in Appendix~\ref{app:crosssections}). The  isospin breaking suppression $\epsilon$ is present 
only in the coupling to pions but not to other hadrons (see Appendix~\ref{app:crosssections} for the unsuppressed Kaon cross section).
Factoring that away and considering the contributions from other states above $T_c$, it is reasonable to expect $\kappa\sim {\cal O}(0.1)$
at around $T_c$ and slightly above. 

Second, recent attempts in pure $SU(3)$ lattice gauge theory estimate the rate for the zero mode 
$\Gamma_{\text{sph}}\simeq \kappa_{\rm latt} T^4$ with $\kappa_{\rm latt}$ ranging from ${\cal O}(0.1)$ to ${\cal O}(0.01)$ as the temperature
increases from $T_c$ to a GeV~\cite{Altenkort:2020axj,Mancha:2022umq}. While the presence of light quarks would strongly suppress the contribution
to the axion rate for the zero mode~\cite{McLerran:1990de,Berghaus:2020ekh}, such a suppression is not expected to occur 
for the typical modes with $E\gtrsim T$.

The corresponding effect for $\Delta N_{\rm eff}$, assuming  an averaged rate with $\kappa=0.1$ or $\kappa=0.01$, turned on only in the interval $T_c<T<2$~GeV (in addition to the momentum dependent production below $T_c$), is represented by the dashed purple curves in fig.~\ref{fig:DeltaNeff}. We stress that our dimensional analysis estimate for the non-perturbative contribution is provided only to highlight 
its potential importance, especially for upcoming experiments.

\section{Current bound and outlook}

Using the ``Pions only" curve in~Fig.~\ref{fig:DeltaNeff} we set a conservative upper bound on $m_a$ from cosmological datasets.
We employ the full Planck 2018 high-$\ell$ and low-$\ell$ TT, TE, EE and lensing likelihoods~\cite{Planck:2019nip}, and additionally Baryon Acoustic Oscillations (BAO) measurements from 6dFGS at $z = 0.106$~\cite{Beutler:2011hx}, from the MGS galaxy sample of SDSS at $z = 0.15$~\cite{Ross:2014qpa},
and from the CMASS and LOWZ galaxy samples
of BOSS DR12 at $z = 0.38, 0.51$, and $0.61$~\cite{BOSS:2016wmc}, including the measurements of the rate of growth of structure~\cite{BOSS:2016ntk}, (we use the ``consensus" likelihood of~\cite{BOSS:2016wmc}), and the Pantheon Supernovae data~\cite{Pan-STARRS1:2017jku}.\footnote{We do not include constraints from Big Bang Nucleosynthesis because of the associated theoretical and observational uncertainties (see~\cite{Planck:2018vyg} for a discussion), nor from Lyman-$\alpha$ forest observations because of the lack of dedicated simulations.} We modified the Boltzmann solver {\tt CLASS}~\cite{Lesgourgues:2011re, Blas:2011rf} to include the axion species and its actual distribution function, rather than using the Bose-Einstein distribution (see Appendix~\ref{sec:cosmo} for details).  We ran a Markov Chain Monte Carlo (MCMC) analysis of the eight-parameter $\Lambda$CDM+$\sum m_\nu$+$m_a$ cosmological model, where $\sum m_\nu\geq 0.06~\text{eV}$~\cite{ParticleDataGroup:2022pth} is the sum of the neutrino masses, using the {\tt MontePython} sampler~\cite{Audren:2012wb, Brinckmann:2018cvx}. We find $m_a\leq 0.24~\text{eV}$, corresponding to $\Delta N_{\text{eff}}\lesssim 0.19$ (red line with star in Fig.~\ref{fig:DeltaNeff}), and $\sum m_\nu\leq 0.14$, both at $95\%$ C.L.; the 1d and 2d posterior distributions of $m_a$ and $\sum m_\nu$ are reported in Fig.~\ref{fig:posterior} (see Appendix~\ref{sec:cosmo} for results on all cosmological parameters). Our result for the axion mass is stronger than previous cosmological bounds in the literature, despite employing more conservative rates, because of the proper inclusion of momentum dependence. In particular, we do not rely on interpolated production rates between LO $\chi$PT
and perturbative QCD amplitudes, such as in~\cite{DEramo:2022nvb}. The use of the actual axion distribution function, rather than a Bose-Einstein distribution with the same energy density, leads to a minor relaxation in the axion mass bound ($\sim 4\%$) with the combination of datasets above.

\begin{figure}[t]
		\includegraphics[width=0.5\textwidth]{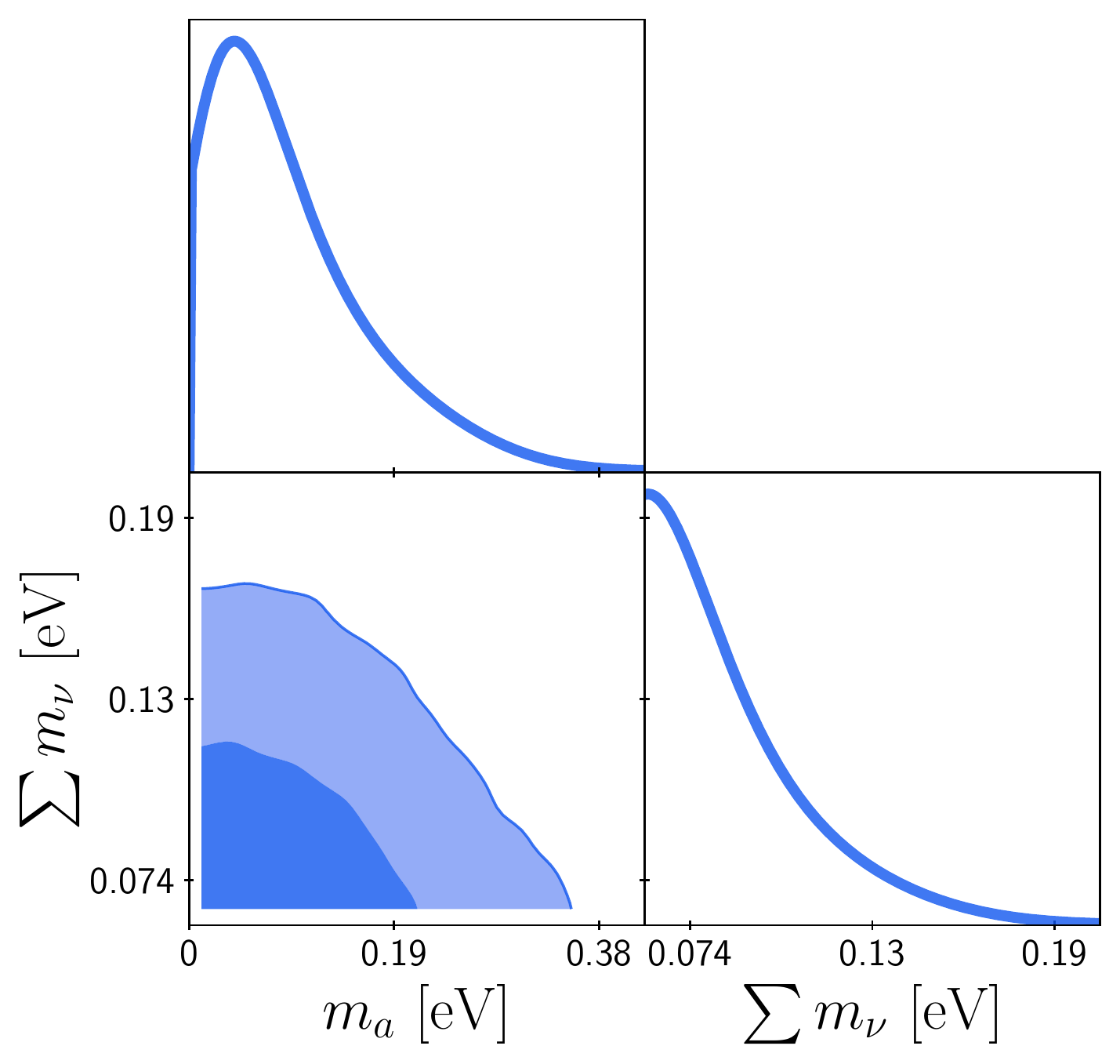}
		\caption{1d and 2d  posterior distributions for $\sum m_\nu$ and $m_a$. The darker/lighter shades correspond to 1$\sigma$ and 2$\sigma$ confidence level regions, extracted from the joint Planck18 TT+TE+EE+low $\ell$+lensing+BAO+Pantheon likelihood.}
		\label{fig:posterior}
	\end{figure}

Our conservative bound can be either weaker or stronger than the astrophysical constraints from globular clusters~\cite{Ayala:2014pea}   (see~\cite{Dolan:2022kul} for a recent more constraining update), depending on the UV structure of the axion model. It is significantly weaker than the constraints of refs.~\cite{Carenza:2020cis, Buschmann:2021juv}, which are however subject to astrophysical uncertainties. In contrast, our bound only relies on standard cosmology below $T_c$.

While we did not include production from the non-perturbative regime in our MCMC search in the conservative spirit of our work, Fig.~\ref{fig:DeltaNeff} shows that it could significantly strengthen the bound already with current datasets. Even more importantly, as shown on the right vertical axis of Fig.~\ref{fig:DeltaNeff}, the upcoming Simons Observatory~\cite{SimonsObservatory:2018koc} and CMB-S4~\cite{CMB-S4:2022ght} will probe axion production during the QCD crossover. In particular, they could reach $m_a\sim 0.01~\text{eV}$, and possibly even below, close to the region where cold axions can be the dark matter (note that the dashed purple curves in Fig.~\ref{fig:DeltaNeff} do not include production above $T=2~\text{GeV}$ nor the enhancement due to the rapid variation of $g_{*, S}$ above $T_c$). 

Such exciting possibilities motivate a dedicated study of axion production rates by non-perturbative methods,
for arbitrary axion momenta, beyond the attempts made so far only for the zero mode. 

\acknowledgments

We thank Ricardo Z.~Ferreira for discussion in the initial stages of this work,~ Miguel Escudero, Vincent Mathieu, Mehrdad Mirbabayi, Juan Torres-Rincon, Marko Simonovic,  Alessandro Strumia, for useful discussions and 
correspondence.

The work of A.N. is supported by the grants PID2019-108122GB-C32 from the Spanish Ministry of Science and Innovation, Unit of Excellence Mar\'ia de Maeztu 2020-2023 of ICCUB (CEX2019-000918-M) and AGAUR2017-SGR-754. A.N. is grateful to the IFPU (SISSA, Trieste) and to the Physics Department of the University of Florence for the hospitality during the course of this work.

\appendix

\section{Boltzmann equations} 
\label{app:boltzmann}

Here we provide more details about the numerical integration of the Boltzmann equation~\eqref{eq:boltzmann}. Rather than integrating over cosmic time $t$, we used the dimensionless variable $x\equiv m_\pi/T$, where the conversion between time and temperature is obtained by imposing entropy conservation, i.e. $d/dt(R^3 g_{*, S} T^3)=0$,  where at weak coupling $g_{*,S}(T)=\sum_{j={\rm bosons}} (\frac{T_j}{T})^3 g_j+\frac{7}{8} \sum_{j={\rm fermions}} (\frac{T_j}{T})^3 g_j$ includes all relativistic particles, with degrees of freedom $g_j$ and possibly with their own temperature $T_j$. This amounts to the replacement $\frac{d}{dt}= (3 x H g_{*,S})/(3 g_{*,S} + T dg_{*,S}/dT) \frac{d}{dx}$. We take $g_{*, S}$ from the numerical lattice results of~\cite{Borsanyi:2016ksw}.

We then solved the resulting equation from $T_{\text{i}}=150~\text{MeV}$ to $T_{\text{f}}=30~\text{MeV}$, for a discrete set of 150 comoving
momenta, uniformly distributed between $|{\bf p}|_{\text{min}}=T_{\text{f}}/5 \cdot R(t_i)$ and $|{\bf p}|_{\text{max}}=30~T_{\text{i}}\cdot R(t_i)$, at the initial time $t_i$, so that physical momenta at any time $ t $ are comprised between $|{\bf k}|_{\text{min}}=|{\bf p}|_{\text{min}}/R(t)$ and $|{\bf k}|_{\text{max}}=|{\bf p}|_{\text{max}}/R(t)$. We computed the energy density in relic axions at $T_f$ using the expression for a massless particle $\rho_a=R^{-4} \int d^3{\bf p}/(2\pi)^3 |{\bf p}| f_{\bf p}$.

Let us now discuss the commonly used momentum-independent approach. Starting from the momentum dependent rates \eqref{eq:axionrate}, one introduces the damping rate 
$\Gamma \equiv {\rm Im} \Pi_a^R/E=\Gamma^>/(1+f^{\rm eq}_{{\bf p}})$, with 
\begin{equation}
    \Pi_a^R\equiv i\int \hspace{-4pt}d^4 x\, e^{ik^\mu x_\mu}\,\left\langle\left[\frac{\alpha_s}{8\pi f_a}G\tilde G(x^\mu),\,\frac{\alpha_s}{8\pi f_a}G\tilde G(0)\right]\Theta(t)\right\rangle,
\end{equation} 
being the retarded axion self energy (see e.g.~\cite{Laine:2016hma} for a review) and $\Theta(t)$ the Heaviside function. Then the Boltzmann equation~\eqref{eq:boltzmann} takes the simpler form
\begin{eqnarray}
\frac{df_{{\bf p}}}{dt} = (f^{\rm eq}_{{\bf p}}- f_{{\bf p}} )\, \Gamma\, , 
\end{eqnarray}
By integrating this equation in $d^3 \mathbf{p}/(2\pi)^3$, assuming $f_{{\bf p}}$ to be of the form (as explained in the main text this assumption is unjustified for the axion case):
\begin{eqnarray}
 f_{{\bf p}}=f^{\rm eq}_{{\bf p}} \, n/n^{\rm eq},
 \label{assumption}
\end{eqnarray}
with $n=R^{-3} \int d^3{\bf p}/(2\pi)^3 f_{\bf p}$, and using the dimensionless variable $x$, one obtains the momentum-independent Boltzmann equation:
\begin{eqnarray}
      \label{standardBoltzmann}
     \frac{dY}{d\log x}  = (Y^{\rm eq}-Y) \frac{\overline{\Gamma}}{H}
     \left( 1- \frac13\frac{d\log g_{*, S}}{d\log x} \right)\, ,
\end{eqnarray}
where $Y\equiv n/{\rm s}$ is the yield
and ${\rm s}=2 \pi^2 g_{*,S} T^3/45$ the entropy density. The final relic energy density in this approach is then given by:
 \begin{equation}
     \rho_a\rvert_{x=x_f}=\left. \frac{\pi ^{14/3} }{30} \left(\frac{{\rm s}\, Y}{\zeta_3}\right)^{4/3}\right. \bigr\rvert_{x=x_f}\, .
     \label{oldrho}
\end{equation}
Such an equation is correct when a given massless species is thermal at high temperature and decouples instantaneously, retaining its Bose-Einstein form after decoupling. Note that, however, this is inconsistent in all other cases (e.g. when starting with zero abundance and never reaching equilibrium), and it is inconsistent with our previous assumption eq.~(\ref{assumption}), since the energy density computed by inserting eq.~(\ref{assumption}) into the integral for $\rho_a$, does not lead to eq.~(\ref{oldrho}).

\begin{figure}[t]
		\includegraphics[width=0.7\columnwidth]{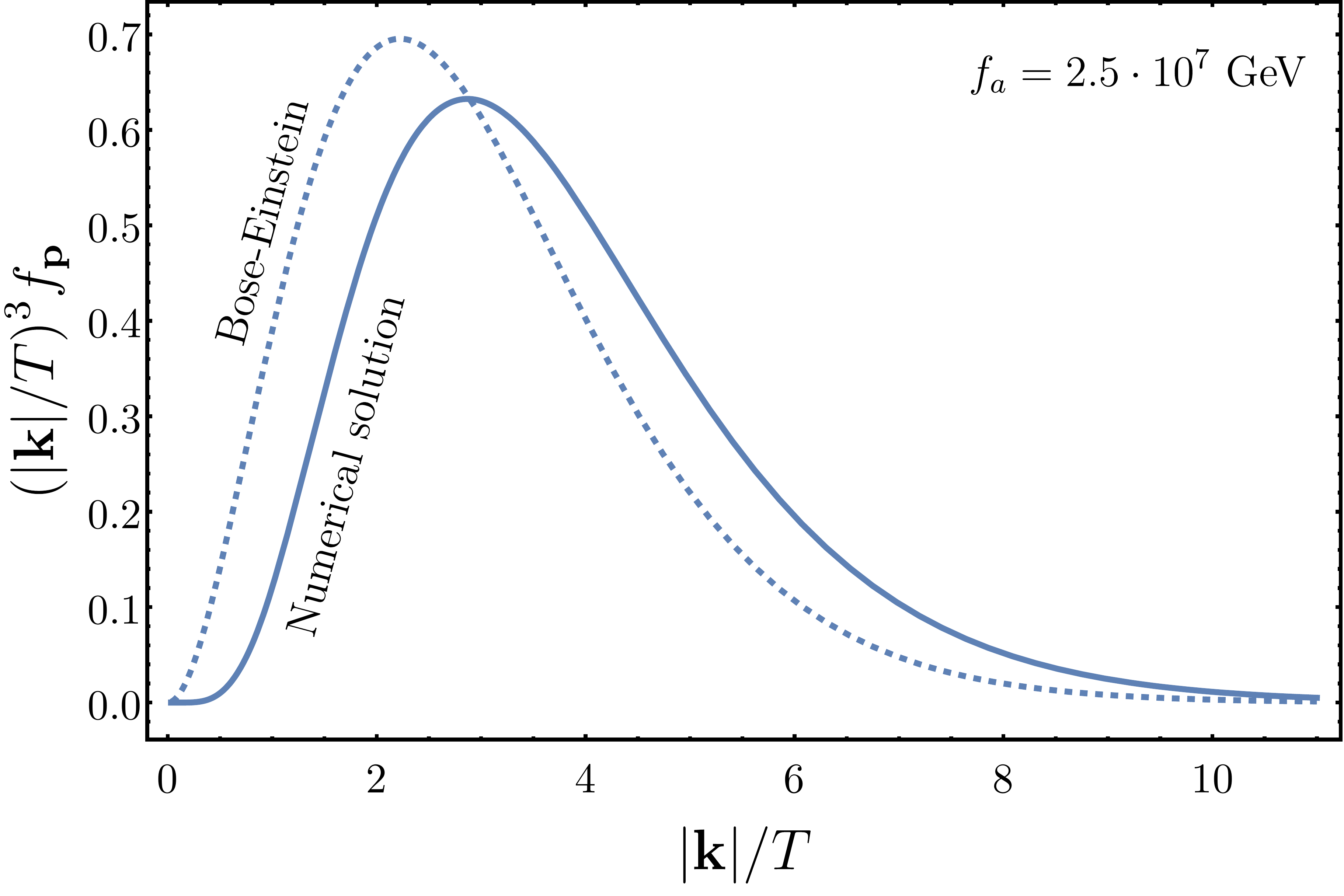}
		\caption{Distortion of the axion spectrum for an example value of $f_a$, as obtained by solving the momentum dependent Boltzmann equations with the rate from scatterings with pions only, below $T_c$ (solid curve). The Bose-Einstein distribution (dashed curve) leads to the same energy density as the numerical solution lead to the same energy density, but the actual distribution is shifted to higher momenta.}
		\label{fig:distortion}
	\end{figure}

One can visualize the spectral distortion of the actual distribution function obtained by solving the full momentum-dependent equation, by comparing with a Bose-Einstein distribution. For the comparison to be meaningful, we set the temperature of the equilibrium distribution such that the resulting energy density is the same as that obtained with the actual distribution. The result is reported in Fig.~\ref{fig:distortion} and shows that that the actual distribution is broader and peaked to higher momenta than the Bose-Einstein distribution.
    
In all cases, either using the exact treatment or the momentum-independent calculation, the relic axion energy density is translated into an effective number of neutrino species at $T_f= \left. 30~\text{MeV}, \Delta N_{\rm eff}\rvert_{T=T_f}\equiv  \frac{8}{7}\frac{\rho_a}{\rho_\gamma}\right.\rvert_{T=T_f}$, then multiplied by $[g_{*, S}(T_f)/g_{*, S}(T_{\nu})]^{-4/3}$, where $T_{\nu}=3~\text{MeV}$ corresponds to the temperature at which neutrinos start decoupling. 
     
For the second analysis presented in this paper, i.e.~including the potential production due to non-perturbative effects during the QCD crossover, we used a hybrid approach, motivated by the lack of a momentum-dependent rate during the QCD crossover. In a first step, we  integrated the momentum independent equation~\eqref{standardBoltzmann} from $T_i=2~\text{GeV}$ to $T_{c}=150~\text{MeV}$, using an averaged rate $\bar{\Gamma}$ determined by dimensional analysis, as explained in the main text. We set initial conditions $Y\rvert_{T=T_i}=Y^{\rm eq}\rvert_{T=1~\text{TeV}}$ and assumed $g_{*, S}=106.75$ at $T\gtrsim \text{TeV}$. The second step of the calculation was made using the momentum-dependent Boltzmann equation, with the corresponding rates for production due to pions only, as explained in section~\ref{pionrate}. Initial conditions for the distribution function at $T_c$ were set according to the would-be equilibrium distributions at $T_{\text{dec}}$, defined as the temperature corresponding to the yield computed with the momentum-independent analysis above $T_c$, i.e. 
\begin{equation}
T_{\text{dec}}\equiv T_c \left. \left(\frac{Y}{Y_{\text{eq}}}\right)^\frac{1}{3}\right.\bigg\rvert_{T=T_c},
\end{equation} 
and $f_{\bf p}|_{T=T_c}=f^{\text{eq}}_{\bf p}\rvert_{T=T_{\text{dec}}}$.

\section{Correspondence between $\pi\pi$ and $a\pi$ scattering}
\label{app:apiscattering}

At leading order in $1/f_a$, the most general effective Lagrangian for
the axion coupled to the lightest two quark fields $q=(u,d)$ can be written as
\begin{equation}
{\cal L}=\bar q\,  \left (i\partial \hspace{-6pt}/+\frac{c_0}{2f_a} \partial \hspace{-6pt}/\, a \,\gamma_5 \right)\, q
-\bar q_L M_a q_R + h.c. \, ,\qquad 
M_a \equiv \left ( 
\begin{array}{cc}
m_u & 0 \\ 0 & m_d
\end{array}
\right)e^{i\frac{a}{2f_a}(1+c_3 \sigma^3) }\,,
\end{equation}
where $q_{R,L}\equiv\frac{1\pm\gamma_5}2 q$
and $c_0$ and $c_3$ are the model dependent axion couplings to the isosinglet and isotriplet axial currents
(e.g. for KSVZ models~\cite{Kim:1979if,Shifman:1979if} $c_3=0$, while for DFSZ models~\cite{Zhitnitsky:1980tq,Dine:1981rt} $c_3=c_u-c_d$, where $c_{u,d}$ are the couplings to the up and down quarks axial currents).

At low energies the effective theory for axion and pions reads:
\begin{equation} \label{eq:LagChPT}
{\cal L}_{\pi}=\frac{f_\pi^2}{4}
{\rm Tr} \left [ \partial_\mu U\partial^\mu U^\dagger + 2 B_0 (M_a U^\dagger + U M_a^\dagger )
\right ] + \dots \qquad U\equiv \exp(i \vec \pi \cdot \vec \sigma /f_\pi)\,, 
\end{equation}
where $\vec \sigma$ are the Pauli matrices, $\vec \pi$ are the pion fields, the constant $B_0= m_\pi^2/(m_u+m_d)$ at leading order and 
dots represent higher order terms in the derivative and quark mass expansion. 
Note that at leading order in the chiral expansion $c_0$ does not appear.
Indeed since at leading order in the quark mass expansion the pion axial current 
is orthogonal to the isosinglet one, the contributions proportional to the coefficient $c_0$ 
will always be suppressed by at least one power of the quark masses, therefore they will
not dominate at high energies.

At leading order the axion only appears in the mass term, which can be written also as
\begin{equation}
\frac{m_\pi^2 f_\pi^2}{2} {\rm Re}\, {\rm Tr}\left [
(1-\epsilon \sigma^3) \left (U e^{-i\frac{a}{2f_a}(1+c_3 \sigma^3)}\right)
\right] \,.
\end{equation}
In this frame there is an $a$-$\pi^0$ mixing that can be rotated away by an orthogonal rotation
with angle $\theta_{a\pi}$:
\begin{equation}
\theta_{a\pi}\simeq \frac{(\epsilon-c_3)f_\pi}{2 f_a}\,,
\end{equation}
up to higher order terms in $f_\pi/f_a$. From this it follows that the diagonalization
to mass eigenstates will introduce extra axion couplings (besides those originating directly from
$M_a$ and thus suppressed by the quark masses) from shifts of the neutral pions:
\begin{equation}
\pi^0= \cos (\theta_{a\pi})\, \pi^0_{\rm phys} + \sin(\theta_{a\pi})\, a_{\rm phys} 
\simeq \pi^0_{\rm phys}+ \theta_{a\pi} a_{\rm phys}\,.
\end{equation}
It follows that, at each order in chiral perturbation theory, the leading
axion couplings at high energies (i.e. those with the maximum number of derivatives,
not suppressed by light quark masses) will be proportional to the corresponding 
$\pi^0$ ones (the proportionality factor being given by the mixing angle itself),
namely:
\begin{equation}
{\cal M}_{a\pi^i \to \pi^j \pi^k} = \theta_{a\pi} \cdot {\cal M}_{\pi^0\pi^i \to \pi^j \pi^k}
+ {\cal O}\left( \frac{m_\pi^2}{s}\right) \,.
\end{equation}

The $m_\pi^2/s$ corrections are the highest at low energy close to threshold.
There the relative error can be estimated using the LO formula $\lvert\mathcal{M}_\text{LO}\rvert^2=\epsilon^2 (s^2+t^2+u^2-3 m_\pi^4)/(4f^2_af^2_\pi)$ and is bounded by 
\begin{equation}
\label{eq:relerror}
\frac{\Delta |\mathcal{M}|^2}{|\mathcal{M}|^2}= \frac{\xi}{1-2\xi}\leq \frac1{11} \,,
\qquad \xi\equiv\frac{3m_\pi^4}{2(s^2+t^2+u^2)}\,,
\end{equation}
where the bound is saturated at $s=4m_\pi^2$, $t=u=-\frac12 m_\pi^2$.
Therefore the asymptotic high energy formula is expected to hold also at low
energies within 10\% accuracy, better than the LO $\chi$PT formula.

\section{Axion Kaon amplitudes}
\label{app:crosssections}

To estimate the importance of axion scattering with heavier hadrons 
below $T_c$ we computed at leading order the scattering amplitude with single kaons ($K^\pm$, $K^0$ and ${\overline K}^0$), which are the next to lightest states after pions. The interactions are described by the same effective Lagrangian of pions in eq.~(\ref{eq:LagChPT}), the only
difference being that the field is now defined as a $3\times 3$ unitary matrix $U=\exp(i\vec \lambda\cdot \vec \Pi)$ (with $\vec \lambda$ the
Gell-Mann matrices and $\vec \Pi$ the eight meson fields including pions and kaons~\cite{Gasser:1983yg}), same for the axion dependent quark mass matrix $M_a={\rm diag}(m_u,m_d,m_s)e^{ia/(3f_a)}$, where $m_s$ is the strange quark mass.

The total amplitude modulus squared 
for the processes $a K \to \pi K$
(which include $a K^\pm \to \pi^0 K^\pm,\, \pi^\pm K^0$; $a K^0\to \pi^0 K^0,\,\pi^-K^+$;
$a \bar K^0\to \pi^0 \bar K^0,\,\pi^+ K^-$) at leading order and in the limit $m_\pi\to 0$ (consistent with the $m_\pi^2/s\to 0$ limit we used
for the pions) turns out to be  
\begin{equation}
|{\cal M}_{a K\to\pi K}|^2=
\frac{3\,t^2}{16 f_\pi^2 f_a^2}(1-\epsilon^2)+
\frac{(s-m_K^2)(s+t-m_K^2)}{2f_\pi^2 f_a^2}\epsilon^2\,,
\end{equation}
where in this case $s\equiv (p_a+p_K)^2$ and $t\equiv (p_a-p_\pi)^2$.

\begin{figure}[t]
		\includegraphics[width=0.8\columnwidth]{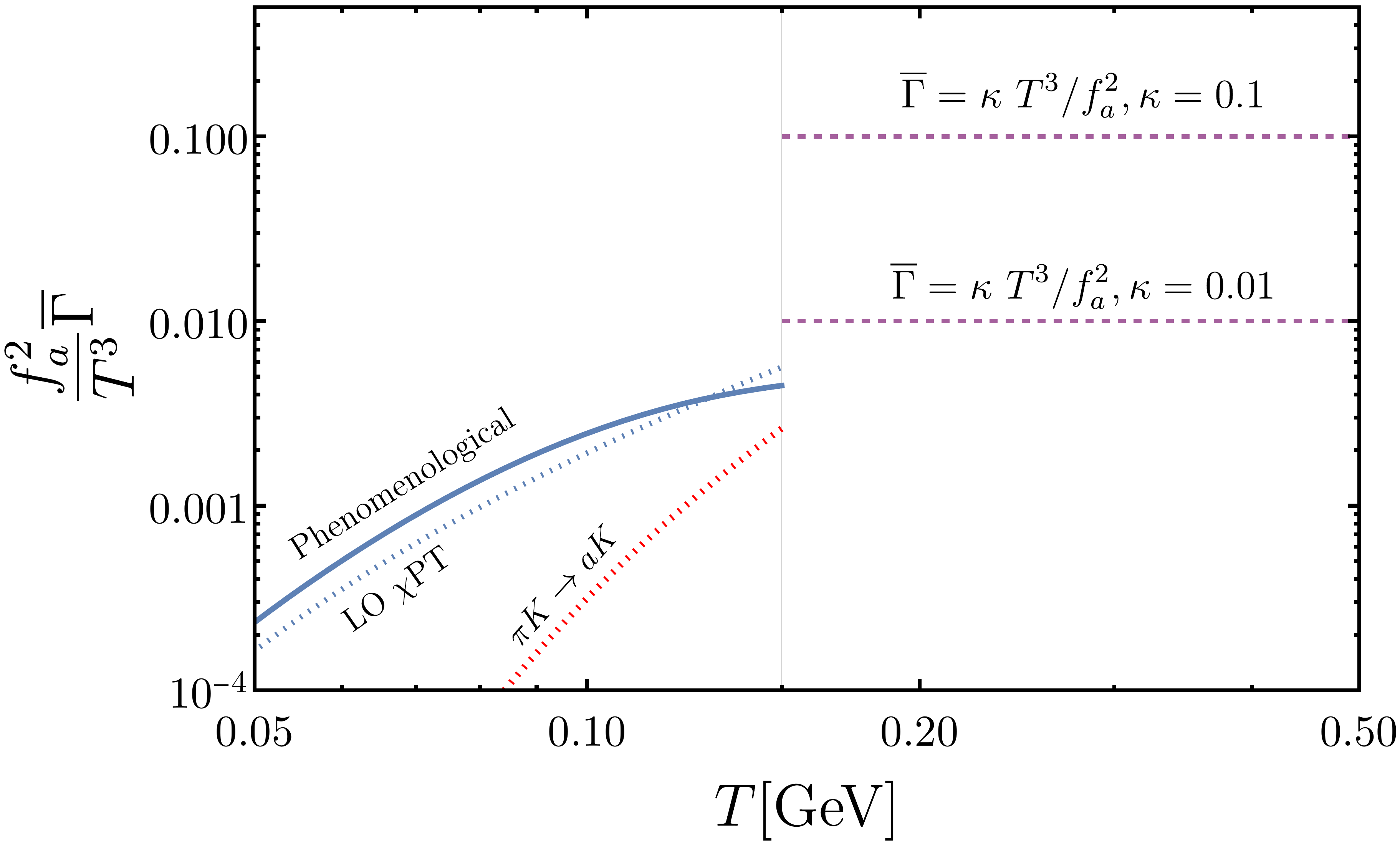}
		\caption{Tentative averaged axion production rate above $T_c$ (see text for description), based on dimensional analysis. The phenomenological and LO rates for $a\pi$ and $a K$ scattering below $T=150~\text{MeV}$ are also shown for comparison.}
		\label{fig:tentrate}
	\end{figure}

Note that the Boltzmann suppression in the scattering with kaons below $T_c$ is partially compensated by the absence of the overall isospin suppression $\epsilon^2$ present for the pions. Indeed the axion rate due to scatterings with kaons would become comparable to the pions one not much after $T_c$, at around $T=200$~MeV (see fig.~\ref{fig:tentrate}).

\section{Computation of Scattering Rates} 
\label{sec:numerics}

Here we provide more details about the numerical calculation of scattering rates. We start from the momentum-dependent rate of the destruction process $a\pi\rightarrow \pi\pi$, given by:

\begin{equation}
\label{eq:Gamma>}
    \Gamma^> = \frac{1}{2 E}\int \left(\prod_{i=1}^3\frac{d^3{\bf k}_{i}}{(2\pi)^3 2E_{i}}\right)f^{\rm eq}_{{1}} (1+f^{\rm eq}_{{2}})  (1+f^{\rm eq}_{{3}}) (2\pi)^4\delta^{(4)}(k^\mu+k^\mu_{1}-k^\mu_{2}-k^\mu_3)|\mathcal{M}|^2,
\end{equation}
where the subscripts $1,2,3$ refer respectively to the incoming pion and the two outgoing pions and $|{\mathcal M}|^2$ is the modulus squared of the scattering amplitude.

For the process of our interest, the integral above can be reduced to an integral over four variables. The steps are as follows, (we follow closely the analogous calculation for neutrino decoupling~\cite{Hannestad:1995rs}): first, one uses the identity $d^3{\bf k}_3/(2E_3)=d^4k_3\,\delta(k^\mu_3 k_{3,\mu}-m_3^2)\Theta(E_3)$ to integrate over $k^\mu_3$. Then one introduces the scattering angles $\alpha, \theta, \beta$ as:
\begin{equation}
    \cos\alpha=\frac{\mathbf{k}\cdot \mathbf{k}_1}{|\mathbf{k}| |\mathbf{k}_1|}, \quad~\cos\theta=\frac{\mathbf{k}\cdot \mathbf{k}_2}{|\mathbf{k}| |\mathbf{k}_2|},\quad \frac{\mathbf{k}_1\cdot \mathbf{k}_2}{|\mathbf{k}_1| |\mathbf{k}_2|}=\cos\alpha\cos\theta + \sin\alpha\sin\theta\cos\beta, \,.
\end{equation}
The remaining integrations are over $d^3 {\bf k}_1=|\mathbf{k}_1|^2 d|\mathbf{k}_1| \, d\cos\alpha \, d\beta, \, d^3 {\bf k}_2=|\mathbf{k}_2|^2 d{|\mathbf{k}|_2} \, d\cos\theta \, d\eta$, where $\eta$ is the polar angle associated to ${\bf k}_2$. For the process under consideration, the integration over $\eta$ is trivial. Writing $k_3^2-m_3^2=h(\beta)$, the leftover $\delta$ function can be used to integrate over $\beta$. In order to do so, one determines the zeros of $h(\beta)$, which is in fact a linear function of $\cos\beta$. Therefore the roots have the same value of $\cos\beta$, and correspondingly two opposite values of $\sin\beta$. Since the integrand only depends on $|\sin\beta|$, the integral can be restricted from $0$ to $\pi$, to select only one value of $\sin\beta$, then multiply by two. Furthermore, the condition $\cos^2\beta\leq 1$ must be imposed. This leads to the following integral:
\begin{align}
\label{eq:momdeprate}
 \nonumber   \Gamma^>&=\frac{1}{2 E}\frac{1}{(2\pi)^4}\int_0^\infty \frac{d|\mathbf{k}_1|}{2E_1}\int_0^\infty \frac{d|\mathbf{k}_2|}{2E_2}\int_{-1}^{1} d\cos\alpha\int_{-1}^{1} d\cos\theta f^{\rm eq}_{{1}} (1+f^{\rm eq}_{{2}})  (1+f^{\rm eq}_{{3}})\\
    &\times \frac{2\Theta(E+E_1-E_2)\Theta(1-\cos\beta_0^2)}{|h^{'}(\beta_0)|}|\mathbf{k}_1|^2 |\mathbf{k}_2|^2|\mathcal{M}|^2,
\end{align}
where $f^{\rm eq}_{{3}}$ is now a function of the integration variables. The modulus squared of the amplitude is a function of the Mandelstam variables $s=(k^\mu+k^\mu_1)^2$ and $t=(k^\mu-k^\mu_3)^2$ ($u$ is eliminated using $s+t+u=3m_\pi^2$), which can be expressed in terms of the integration variables above. 

The LO $\chi$PT amplitude for $a\pi\rightarrow\pi\pi$ is given in Appendix~\ref{app:apiscattering}, above \eqref{eq:relerror}. In order to obtain an amplitude from the phenomenological partial wave phases of $\pi\pi$ scattering~\cite{Pelaez:2004vs, Garcia-Martin:2011iqs}, the amplitude is decomposed in definite isospin $I=0,1,2$ in the $s$-channel (see e.g.~\cite{Gasser:1983yg}):
\begin{align}
    T^0(s,t) &= 3A(s,t,u)+A(t,u,s)+A(u,s,t),\\
    T^1(s,t) &= A(t,u,s) - A(u,s,t),\\
    T^2(s,t) &= A(t,u,s) + A(u,s,t),
\end{align}
and then each amplitude is expanded into partial wave amplitudes $t^I_l$, according to:
\begin{equation}
    T^I(s,t)=32\pi\sum_{l=0}^{\infty} (2l+1)P_l(\cos\xi)t^I_l(s),
\end{equation}
where $t=(s-4m^2_\pi)(\cos\xi-1)/2$ and $P_l$ is the $l$-th legendre polynomial. Finally, the experimental data are reported in terms of real phase shifts $\delta^I_l$, related to the amplitudes by:
\begin{equation}
    t^I_l(s)=\sqrt{\frac{s}{s-4m^2_\pi}}\frac{e^{2i\delta^I_l(s)}-1}{2i}=\sqrt{\frac{s}{s-4m^2_\pi}}\frac{1}{\cot\delta^I_l(s)-i}.
\end{equation}
We took the phenomenological parametrizations for the $\delta^I_l(s)$ (specifically the $\delta^0_0, \delta^1_1$ and $\delta^2_0$, we checked that the other waves give a negligible contribution to the rates) from Appendix A of~\cite{Garcia-Martin:2011iqs}, restricting ourselves to $\sqrt{s}\leq 2m_{K}$, where elastic waves can be used. We used two different phenomenological parametrizations for each of the two $\delta^0_0$ and $\delta^2_0$ shifts: one until $\sqrt{s}=850~\text{MeV}$ and a second one in the region  $850~\text{MeV}\leq\sqrt{s}\leq 2m_K$, following the analysis of~\cite{Garcia-Martin:2011iqs}. The modulus squared of the total amplitude is then given by:
\begin{equation}
\label{eq:mpipi}
    |\mathcal{M}|_{\pi\pi}^2=\frac{1}{2}\frac{32\pi^2}{3}\left(|t^0_0|^2+3|3 t^1_1\cos\xi|^2+5|t^2_0|^2\right).
\end{equation}
As discussed in the main text, the amplitude modulus squared above can be related to that of $a$-$\pi$ scattering via rescaling by $\epsilon^2f_\pi^2/(4 f_a^2)$. 

We performed the four-dimensional integral in \eqref{eq:momdeprate} numerically, with the amplitudes discussed above, by using the \href{https://vegas.readthedocs.io/en/latest/tutorial.html}{\tt vegas} Monte Carlo integrator in Python, based on~\cite{Lepage:2020tgj}. Specifically, we computed $\Gamma^>$ for a discrete set of values of $\log_{10} (|{\bf k}|/m_\pi)=[-3,2]$ and $\log_{10} (m_\pi/T)=[-0.3, 1]$, with steps of $0.1$ in both variables. We then built an interpolating function of two variables using {\tt Mathematica}~\cite{Mathematica}.

We also numerically computed the rate $\overline{\Gamma}$, according to \eqref{eq:totrate} and \eqref{gammas}:
\begin{equation}
    \overline{\Gamma} = \frac{1}{n_{\text{eq}}}\int \frac{d^3{\bf k}}{(2\pi)^3} \frac{f^{\rm eq}_{{\bf p}}}{1+f^{\rm eq}_{{\bf p}}}\Gamma^>.
\end{equation}
Note that this rate differs from the commonly used momentum-independent rate by a factor $1/(1+f^{\rm eq}_{\bf p})$ in the integrand. In practice, this difference is small. We followed the same procedure for the LO $\chi$PT rate for $a K\rightarrow \pi K$.

\section{Rate from strong sphalerons}
\label{app:sphaleron}
Here we evaluate the scale at which non-perturbative collective effects
in the damping rate start being competitive with the perturbative ones
in the simplified case of a pure $SU(N_c)$ theory without fermions.

A known source of such effects are ``strong'' sphalerons that for sufficiently weak coupling have been assessed to have a rate $\Gamma_{\rm sphal}\simeq (N_c \alpha_s)^5 T^4$ \cite{Moore:2010jd}. Such a rate should be identified with what we called $\Gamma^>_{\rm top}(k^\mu=0)$ in eq.~(\ref{eq:axionrate}).  Sphalerons are known to have a size of order $1/(N_c \alpha_s T)$, therefore we expect $\Gamma^>_{\rm top}(E=|\mathbf{k}|<|\mathbf{k}_s|)\simeq \Gamma_{\rm sphal}$ with $|\mathbf{k}_s| \sim  N_c \alpha_s T$. Using eqs.~(\ref{eq:axionrate}) and (\ref{eq:totrate}) we get
\begin{equation} \label{eq:GammaNP}
\overline{\Gamma}_{\rm sphal}=\frac{1}{n^{\rm eq}}\hspace{-8pt}\stackrel{|\mathbf{k}|<|\mathbf{k}_s|}{\int}\hspace{-8pt} \frac{d^3{\bf k}}{(2\pi)^3 2E}\frac{\Gamma_{\rm sphal}}{f_a^2}e^{-E/T} =\frac{(N_c \alpha_s)^5 T^3}{4\zeta_3 f_a^2}
\left(1-\left(1+\frac{|\mathbf{k}_s|}{T}\right)e^{-|\mathbf{k}_s|/T}\right)\,.
\end{equation}

For $N_c=3$ such a rate can be compared with the perturbative estimates performed in QCD in refs.~\cite{Masso:2002np,Graf:2010tv,Salvio:2013iaa,DEramo:2021lgb},
which can be summarized as
\begin{equation}
    \overline\Gamma_{\rm pert}=\frac{\alpha_s^2 T^3}{4\pi^3 f_a^2}F_3
\end{equation}
where $F_3=g_s^2\log(3/(2g_s))^2$ in ref.~\cite{Graf:2010tv} and some more non-trivial function in refs.~\cite{Salvio:2013iaa,DEramo:2021lgb}.
\begin{figure}[t]
    \centering
    \includegraphics[width=0.7\textwidth]{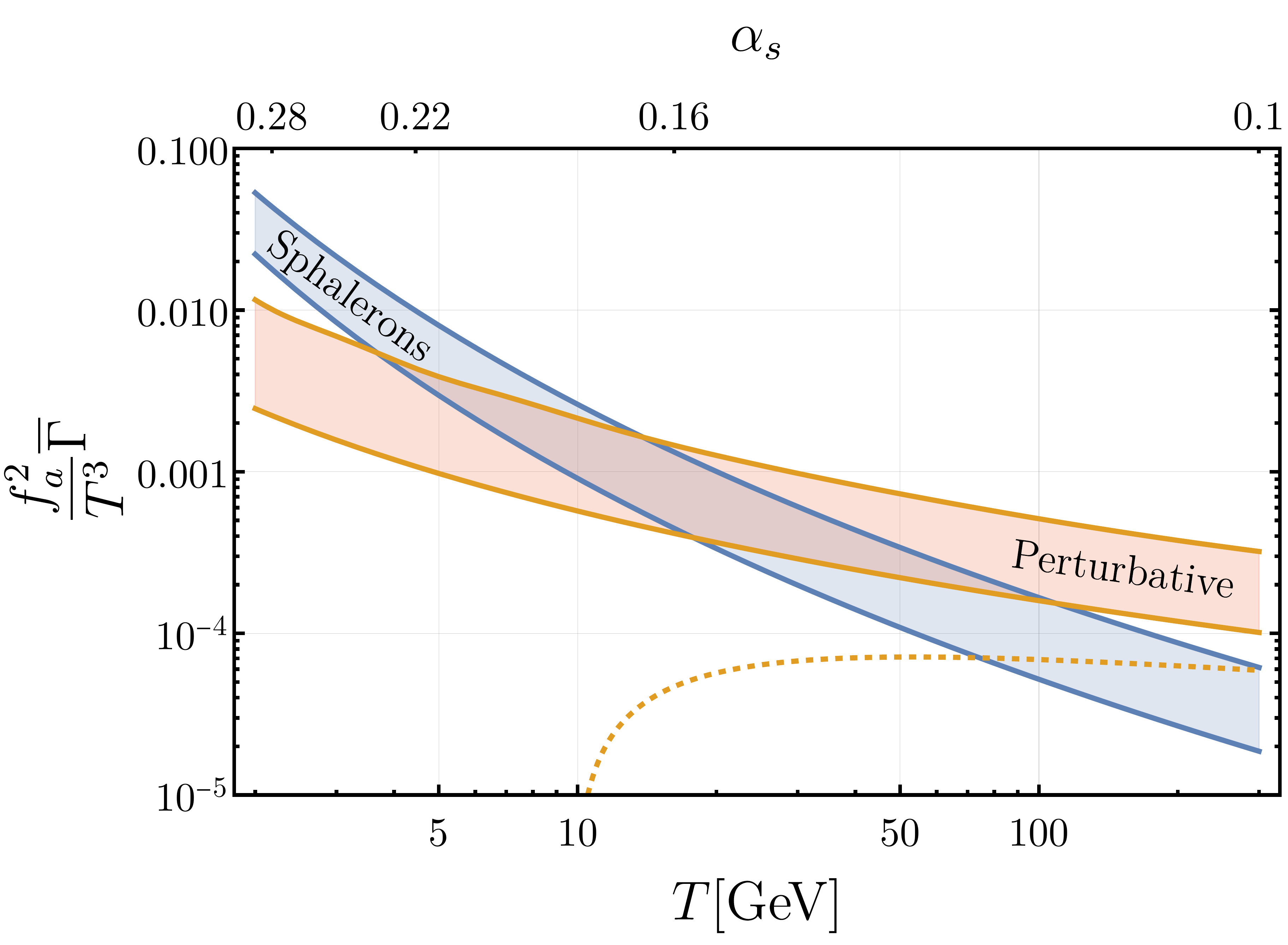}
    \caption{Comparison between non perturbative and perturbative contributions
    to axion rates as a function of temperature, see text for more details.}
    \label{fig:GammaNPvsP}
\end{figure}

In fig.~\ref{fig:GammaNPvsP} we compare $\overline \Gamma_{\rm sphal}$
for the two choices $|\mathbf{k}_s|=3\alpha_s T$ (lower blue line) and $|\mathbf{k}_s|=6\alpha_s T$ (upper blue line), with $\overline \Gamma_{\rm pert}$
for the three choices: $F_3$ from ref.~\cite{Graf:2010tv} (orange dashed), $F_3=g_s^2$ (lower orange curve), $F_3$ from ref.~\cite{DEramo:2021lgb} (upper orange curve). Notice that the choice from ref.~\cite{Graf:2010tv}
breaks down already at moderate values of the coupling, due to the infrared logarithmic effects running negative. The choice $F_3=g_s^2$
represent the expected size of the naive perturbative computation if the logarithmic factor saturates at unity. The last choice instead
implement some partial resummation enhancing the rate.

In any case it appears that collective sphaleron-like effects start
becoming important between $T\sim 100$~GeV and $T\sim 5$~GeV, anyway
much before the breaking of QCD perturbation theory in vacuum. 

The inclusion of (light) fermions in the computation of the non-perturbative effects is technically more challenging\footnote{The effects of sphalerons are known to be 
damped by massless fermions at $|\mathbf{k}|=0$, however it is easy to check
that this is no longer true at $|\mathbf{k}|>0$, in particular 
for the values relevant for us.},
but we do not expect that their presence could sensibly improve the reliability of the pertubative computation.

\section{Cosmological analysis}
\label{sec:cosmo}
Here we outline our implementation of hot axions into the Boltzmann solver {\tt CLASS}, as well as provide extended results of our MCMC analysis.

We implemented the axion as an additional non-cold dark matter degree of freedom ({\tt ncdm}) in {\tt CLASS}, including its spectral distribution obtained by solving the momentum-dependent Boltzmann equation for the ``Pions only" calculation. Specifically, we added a new distribution to the {\tt background.c} file. We found that the integrand of the axion energy density, i.e. $|{\bf k}|^3~f_{\mathbf{p}}$, is well approximated by the following functional form (in particular in the most important region $|{\bf k}|\gtrsim T$, see Fig.~\ref{fig:distortion}), in the range of axion masses of interest:
\begin{equation}
\label{eq:distapprox}
|{\bf k}|^3 f_{\text{approx},\mathbf{p}}= |{\bf k}|^3 \frac{A}{e^{\frac{|{\bf k}|}{bT}+\mu}+1},
\end{equation}
where $A, b$ and $\mu$ are parameters that vary with the axion mass. Their behavior for $m_a\leq 0.38~\text{eV}$ can be simply described by polynomial fits. Notice that the actual distribution function does not grow at small $|{\bf k}|$, in contrast to a standard Bose-Einstein distribution. This different behavior is also visible in Fig.~\ref{fig:distortion}. This feature at small $|{\bf k}|$ is due to the fact that low momenta are not significantly produced for the values of $f_a$ of interest, and that our initial conditions are set to be vanishing at $T_c$. This is the reason why the functional form \eqref{eq:distapprox}, with $+1$ in the denominator, fits the actual distribution function better than the analogous one with $-1$ in the denominator. 

We have thus implemented the axion distribution function as  $f_{\text{approx},\mathbf{p}}$. Notice that this is a function of $q\equiv |{\bf k}|/T$ and $m_a$ only. Additionally, the {\tt ncdm} module requires to specify a temperature parameter $T_{\text{ncdm}}$, which for non-equilibrium distributions is just a normalization parameter. Accordingly, we have set it to be equal to the temperature of one dark species in units of the photon temperature, i.e. $(4/11)^{1/3}$. In practice, since we obtain our distribution function at $T_f = 30~\text{MeV}$ rather than at the temperature of neutrino decoupling $T\approx 3~\text{MeV}$, we have set $T_{\text{ncdm, a}}=(4/11)^{1/3}[g_{*,S}(3~\text{MeV})/g_{*,S}(30~\text{MeV})]^{1/3}$.

The output value of $\Delta N_{\text{eff}}$ obtained with our modified version of {\tt CLASS} matches the result of Fig.~\ref{fig:DeltaNeff} to within $\lesssim 1~\%$ for $0.05~\text{eV}\leq m_a\leq 0.38$. For smaller axion masses the value of $\Delta N_{\text{eff}}$ is anyway negligible, see Fig.~\ref{fig:DeltaNeff}.

For $m_a\geq 0.38$, i.e. in the axion mass range which is anyway strongly constrained by CMB data according to our Fig.~\ref{fig:DeltaNeff}, we neglected the spectral distortion in {\tt CLASS} and set the axion to have a Bose-Einstein distribution, with temperature given  by:
\begin{equation}
T_{\text{ncdm}, a}(m_a) = T_\nu\left[\frac{7}{4}\Delta N_{\text{eff}}(m_a)\right]^{\frac{1}{4}},
\end{equation}
where $T_\nu=(4/11)^{1/3}~\text{eV}$ is the temperature of one neutrino species and we used our result for ``Pions only" in Fig.~\ref{fig:DeltaNeff} for $\Delta N_{\text{eff}}(m_a)$. We again found an analytical function that fits our result for $\Delta N_{\text{eff}}$ well and implemented it in {\tt CLASS}.

For the neutrino sector, we followed the Planck collaboration convention of using three massive neutrinos with degenerate masses, and let their sum free to vary in our MCMC analysis. We set a prior $m_\nu\geq 0.02~\text{eV}$ to take into account the lower bound on the sum of neutrino masses $\sum_\nu m_\nu\geq 0.06~\text{eV}$ from neutrino oscillations~\cite{ParticleDataGroup:2022pth}. 

\begin{table*}[t]
\begin{tabular} {| l | c| }
\hline\hline
 \multicolumn{1}{|c|}{ Parameter} &  \multicolumn{1}{|c|}{~~~Planck18+BAO+Pantheon~~~} \\
\hline\hline
$100 \omega_b              $ & $(2.252)~2.246_{-0.015}^{+0.014}$ \\
$\omega_{cdm }             $ & $(0.119)~0.120_{-0.002}^{+0.001}$ \\
$100~\theta_s              $ & $(1.042)~1.042_{-0.0003}^{+0.0003}$\\
$\ln 10^{10}A_s            $ & $(3.051)~3.054_{-0.016}^{+0.015}$ \\
$n_{s}                    $ & $(0.968)~0.968_{-0.005}^{+0.004}$ \\
$\tau_{reio }              $ & $(0.059)~0.058_{-0.008}^{+0.007}$ \\
\hline
$\sum m_{\nu}                   $ &$\leq 0.14~(95\%~\text{C.L.})$\\
$m_{a}                   $ &$\leq 0.24~(95\%~\text{C.L.})$\\
\hline
$H_0[\text{km/s/Mpc}]$               &
$(68.17)~67.75_{-0.54}^{+0.51}$ \\
$S_8$               &
$(0.819)~0.821_{-0.011}^{+0.011}$ \\
\hline
\end{tabular}
  \caption{The mean (best-fit in parenthesis) $\pm1\sigma$ error of the cosmological parameters obtained by fitting the $\Lambda$CDM + $\sum m_\nu$+$m_a$ model to the cosmological datasets Planck18+BAO. Upper bounds are presented at 95\% CL. $H_0$ and $S_8$ are derived parameters from the eight cosmological parameters.}
  \label{table:results}
\end{table*}

\begin{figure}[]
		\includegraphics[width=\textwidth]{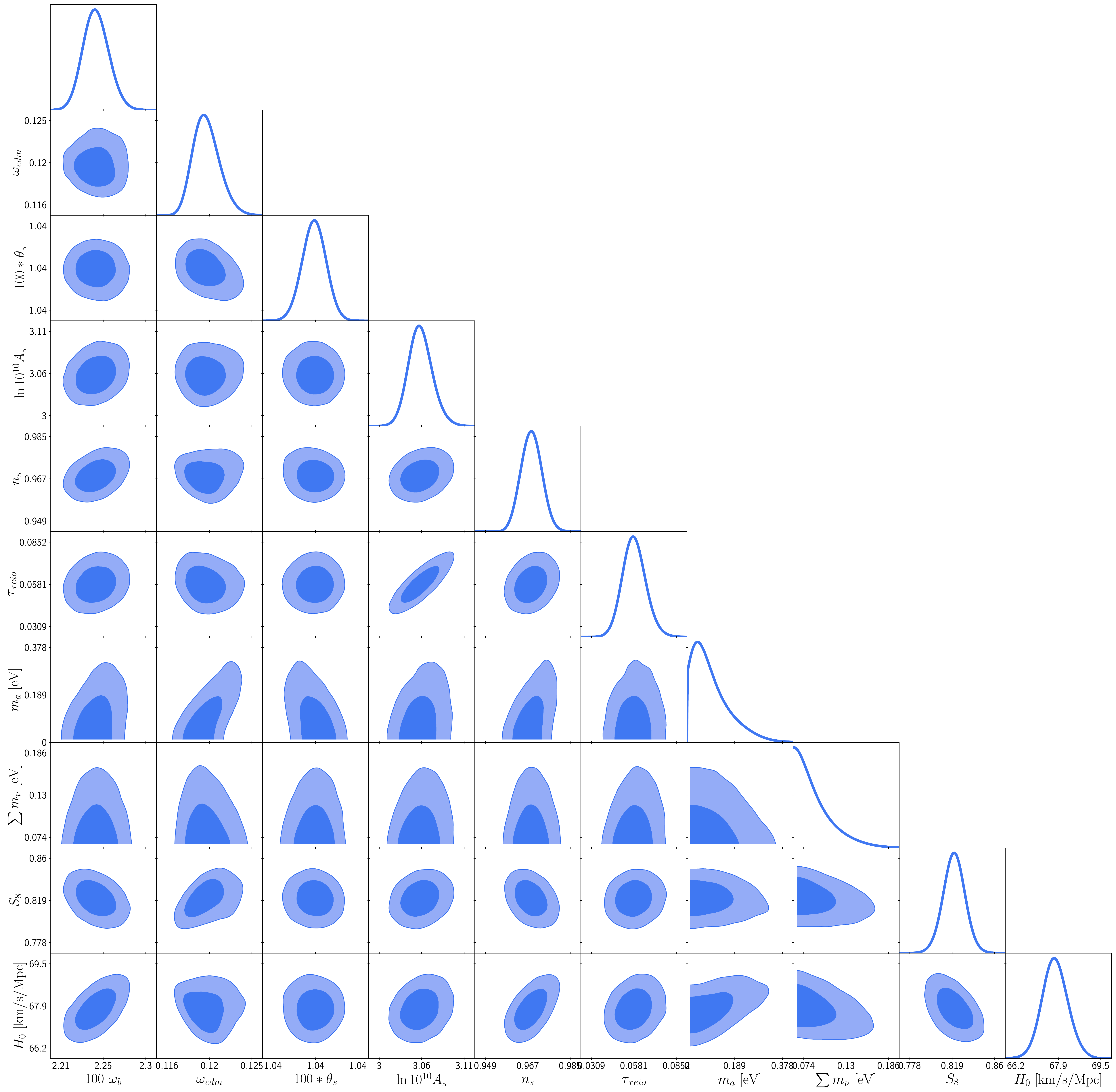}
		\caption{1d and 2d posterior distributions for the $\Lambda$CDM+$\sum m_\nu$+$m_a$ model, extracted from the joint Planck18 TT+TE+EE+low $\ell$+lensing+BAO+Pantheon likelihood.}
		\label{fig:posteriors}
	\end{figure}

We thus obtain a cosmological model with the six standard $\Lambda$CDM parameters plus the additional axion mass $m_a\geq 0$ and the sum of neutrino masses $\sum_\nu m_\nu\geq 0.06~\text{eV}$. We perform our MCMC analysis using {\tt MontePython}, with the full Planck18 TT+TE+EE+low $\ell$+lensing likelihood, the BAO likelihoods {\tt bao\_fs\_boss\_dr12} and {\tt bao\_smallz\_2014}, as well as the {\tt Pantheon} likelihood. We run ten chains, achieving Gelman-Rubin factors $R-1\leq 0.005$ for all parameters. The resulting one- and two-dimensional posterior distributions for all parameters are shown in Fig.~\ref{fig:posteriors} (including $H_0$ and $S_8\equiv \sigma_8\sqrt{\Omega_m/0.3}$. The mean $\pm1\sigma$ error values for the $\Lambda$CDM parameters, together with the $95\%$ C.L. bound for $m_a$, are reported in Table~\ref{table:results}. Additionally, we have checked that our upper bound on the axion mass is unaltered if we use the base $\Lambda$CDM model with only one massive neutrinos and $m_\nu=0.06~\text{eV}$.

\newpage

\bibliography{biblio}

\end{document}